\documentclass[a4paper,11pt]{article}

\usepackage{amsfonts}
\usepackage{amsmath}
\usepackage{epsf}
\usepackage{texdraw}

\newcounter{stuff}

\newtheorem{prop}[stuff]{Proposition}

\newtheorem{thm}[stuff]{Theorem}
\newtheorem{defi}[stuff]{Definition}

\newtheorem{lem}[stuff]{Lemma}

\def\ket#1{\,\vert#1\/\rangle}              

\newcommand{\R}{\mathbb{R}}
\newcommand{\N}{\mathbb{N}}

\renewcommand{\Im}{\mathrm{Im}\,}
\renewcommand{\Re}{\mathrm{Re}\,}

\numberwithin{equation}{section}
\numberwithin{stuff}{section}

\begin{document}

\title{Semiclassical Limits of Extended Racah Coefficients} \author{Stefan
Davids\thanks{e-mail: \texttt{smd}@\texttt{maths.nott.ac.uk}}\\ \\ Department
of Mathematics\\University of Nottingham\\ Nottingham NG7 2RD\\UK} 

\maketitle

\begin{abstract} 

We explore the geometry and asymptotics of extended Racah coeffecients. The
extension is shown to have a simple relationship to the Racah coefficients for
the positive discrete unitary representation series of SU(1,1) which is
explicitly defined. Moreover, it is found that this extension may be
geometrically identified with two types of Lorentzian tetrahedra for which all
the faces are timelike.

The asymptotic formulae derived for the extension are found to have a similar
form to the standard Ponzano-Regge asymptotic formulae for the SU(2) 6j symbol
and so should be viable for use in a state sum for three dimensional Lorentzian
quantum gravity.

\end{abstract}

\newpage

\section{Introduction} \label{intro}

It is widely believed that the Ponzano-Regge state sum model for the group
SU(2) \cite{PR} is equivalent to (2+1) Euclidean quantum gravity with zero
cosmological constant\cite{Hass81}. The state sum is defined in terms of the 6j
symbols, or Racah coefficients, of SU(2), one for each tetrahedron of the
triangulation. The equivalence to quantum gravity arises when one recovers the
exponential of the Regge action  of a tetrahedron from each 6j symbol in a 
suitable asymptotic limit.  Thus we have the equivalence principle discussed in
\cite{Ba95}.

There is a deeply unsatisfactory property of this state sum arising from the 
use of SU(2). This, as the double cover of SO(3), indicates a Euclidean theory.
As a possible consequence we find an inversion of the usual relationship
between the Euclidean and Lorentzian actions; thus the Euclidean tetrahedra
have an oscillatory action, while Lorentzian tetrahedra exponentially decay. 

In this paper we shall attempt to remedy this by a different choice of group;
namely SU(1,1) since this is the double cover of the three dimensional Lorentz
group  SO(2,1). In section \ref{ext} an extension to the SU(2) 6j symbol is
explicitly  defined using the extended symmetries of \cite{Pon74_6} and shown
to satisfy the orthogonality relation. The definition of the SU(1,1) 6j symbol
is given in section \ref{rac} for the positive discrete unitary series,  by
analogy to the SU(2) case, as well as an explicit formula. It is shown to have
a very simple relationship to the extension defined in section \ref{ext} and to
satisfy a Biedenharn-Elliot type relation.

In section \ref{geom} the geometry determined by the extension of the 6j symbol
(or, in view of the results of section \ref{rac}, the SU(1,1) 6j symbol) is
explored in some detail. They are found to correspond to Lorentzian tetrahedra
with timelike faces; the edges may be all spacelike or all timelike depending
on the sign of the Cayley determinant. Finally, in section \ref{asymp}, the
asymptotics of the exentension, corresponding to both types of tetrahedra, are 
derived using the Ponzano-Regge\cite{PR}  formula and the results of sections
\ref{ext} and \ref{geom}. The two asymptotic formulae are found to have a
similar form to the known SU(2) asymptotic formulae found by Ponzano and Regge
and so may be interpreted as a probability arising from a path integral for
three dimensional Lorentzian quantum gravity.

It is intended to pursue these ideas in a future work and develop a full
state sum model for three dimensional Lorentzian quantum gravity.

\section{Extensions of 6j Symbols} \label{ext}

In \cite{Pon74_3} and \cite{Pon74_6} the symmetries of 3j and 6j coefficients
were extended beyond the usual symmetries, which respect the triangle
inequality, to a new domain, which satisfies an anti-triangle inequality. The
extension of the 6j symbol is related to the 6j symbol for the positive discrete 
unitary representation series of SU(1,1). 

To be more precise, the extension of the 3j symbol discussed in \cite{Pon74_3}
corresponds, within a phase, to the explicited calculated 3j symbol for the
coupling of two elements of the discrete series of SU(1,1) given in \cite{HoBi66}. 
For the 6j symbol, the regions associated with the extension to anti-triangle
inequalities, discussed in \cite{Pon74_6}, have been conjectured to be 
related to the 6j symbol for
the discrete unitary representation series of SU(1,1). 
The precise relationship will be derived in
section \ref{rac}.

In this section we shall explicitly compute a transformation of the 6j
symbol to the region conjectured to be associated to these discrete unitary 
representations using the symmetries in \cite{Pon74_6}. We start with some
definitions.

\begin{defi}
We shall use
the symbol $\left| \begin{array}{ccc}
a & b & c \\ d & e & f \end{array} \right|_{SU(2)}$ to denote an ordered
set of real numbers numbers for
which the ordered sets of real numbers
$|abc|_{SU(2)}$, $|cde|_{SU(2)}$, $|afe|_{SU(2)}$ and $|bdf|_{SU(2)}$
each satisfy mutual
triangle inequalities (that is $\pm a\pm b\pm c\ge 0$ where two plus
signs are chosen). 
We shall use the symbol $\left| \begin{array}{ccc}
a & b & c \\ d & e & f \end{array} \right|_{SU(1,1)}$ in a similar way, but here
$|abc|_{SU(1,1)}$, etc. satisfy $c\ge a+b+1$, $a\le b+c$ and $b\le
a+c$ instead of mutual triangle inequalities. Both will satisfy 
the sum of the three elements being at least -1.\footnote{
For the symbols $|abc|_{SU(2)}$, etc this last condition is redundant since one
can show that the mutual triangle inequalities imply the nonnegativity of
$a$, $b$ and $c$}
\end{defi}

\begin{defi}
The 6j symbol defines a map 
$$
\R^6\to\R
$$
given by
$$
\left| \begin{array}{ccc}
a & b & c \\ d & e & f \end{array} \right|_{SU(2)}
\mapsto
\left\{ \begin{array}{ccc}
a & b & c \\ d & e & f \end{array} \right\}_{SU(2)}
$$
while what we shall call the \emph{extension} defines another map $\R^6\to\R$
given by
$$
\left| \begin{array}{ccc}
a & b & c \\ d & e & f \end{array} \right|_{SU(1,1)}\mapsto
\left\{ \begin{array}{ccc}
a & b & c \\ d & e & f \end{array} \right\}_{ext}
$$

The details of these two maps will be given later. 
\end{defi}

\begin{defi} \label{S}
Define a map $S:\R^6 \to\R^6$ 
\begin{align}
a & = \frac{1}{2}\left(a^\prime+b^\prime-d^\prime+e^\prime\right) \label{b1} \\
b & = \frac{1}{2}\left(-a^\prime-b^\prime-d^\prime+e^\prime\right)-1\label{b2} \\
c & = c^\prime\label{b3} \\
d & = \frac{1}{2}\left(-a^\prime+b^\prime+d^\prime+e^\prime\right) \label{b4} \\
e & = \frac{1}{2}\left(a^\prime-b^\prime+d^\prime+e^\prime\right) \label{b5} \\
f & = f^\prime\label{b6}
\end{align}
It should be noted that if one shifts all
the values of the variables by $+\frac{1}{2}$ then this transformation is an
\emph{orthogonal} linear map. The inverses to
equations \ref{b1} - \ref{b6} are
\begin{align}
a^\prime & = \frac{1}{2}\left(a-b-d+e-1\right) \label{a} \\
b^\prime & = \frac{1}{2}\left(a-b+d-e-1\right) \label{b} \\
c^\prime & = c  \label{c} \\
d^\prime & = \frac{1}{2}\left(-a-b+d+e-1\right) \label{d} \\
e^\prime & = \frac{1}{2}\left(a+b+d+e+1\right) \label{e} \\
f^\prime & = f  \label{f}
\end{align}
\end{defi}

\begin{prop}
For $S$ defined in definition \ref{S} we have
\begin{equation}
S:\left| \begin{array}{ccc}
a^\prime & b^\prime & c^\prime \\ d^\prime & e^\prime & f^\prime
\end{array} \right|_{SU(1,1)} \to
\left| \begin{array}{ccc}
a & b & c \\ d & e & f \end{array} \right|_{SU(2)}
\end{equation}
\end{prop}

To prove this, consider the map acting on the ordered sets
$|abc|_{SU(2)}$ associated to $\left| \begin{array}{ccc}
a & b & c \\ d & e & f \end{array} 
\right|_{SU(2)}$. We find
\begin{align}
a+b-c & = e^\prime-d^\prime-c^\prime-1\label{eq1}\\
a-b+c & = a^\prime+b^\prime+c^\prime+1\\
-a+b+c & = -a^\prime-b^\prime+c^\prime-1\\
a+b+c+1 &= e^\prime-d^\prime+c^\prime
\end{align} \begin{align}
c+d-e & = c^\prime+b^\prime-a^\prime\\
c-d+e & = a^\prime-b^\prime+c^\prime\\
-c+d+e & = d^\prime+e^\prime-c^\prime\\
c+d+e+1 & = e^\prime+d^\prime+c^\prime+1\label{eq8}
\end{align} 

\begin{figure}[htb]
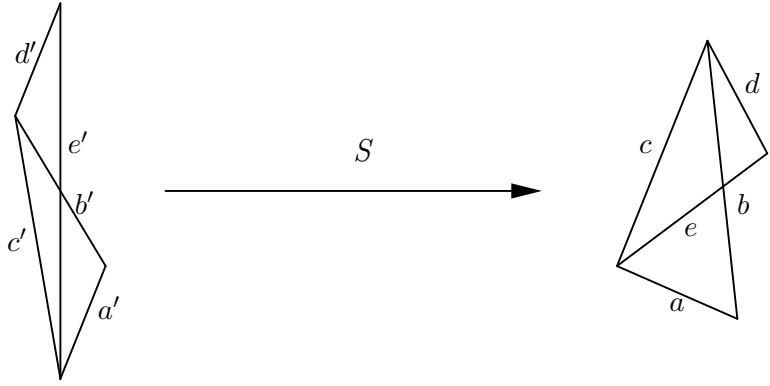

\begin{center}
\begin{texdraw}
\drawdim{mm}
\lvec(0 50) \rmove(1 -20) \htext{$e^\prime$} \rmove(-1 20) 
\rlvec(-6 -15) \rmove(0 7) \htext{$d^\prime$} \rmove(0 -7)
\rlvec(6 -35) \rmove(-7 17) \htext{$c^\prime$} \rmove(7 -17) 
\rlvec(6 15) \rmove(-1 -7) \htext{$a^\prime$} \rmove(1 7)
\rlvec(-12 20) \rmove(8 -13) \htext{$b^\prime$} \rmove(-8 13)

\rmove(20 -10)
\arrowheadtype t:F 
\ravec(50 0) \rmove(-50 0) 
\rmove(25 4) \htext{\textit{\large S}} \rmove(-25 -4)
\rmove(-20 10)

\rmove(80 -20)
\rlvec(20 15) \rmove(-11 -11) \htext{$e$} \rmove(11 11)
\rlvec(-8 15) \rmove(5 -7) \htext{$d$} \rmove(-5 7)
\rlvec(-12 -30) \rmove(3 15) \htext{$c$} \rmove(-3 -15)
\rlvec(16 -7) \rmove(-9 1) \htext{$a$} \rmove(9 -1)
\rlvec(-4 37) \rmove(4 -23) \htext{$b$} 

\end{texdraw}
\end{center}
\caption{A graphic representation of equations \ref{eq1} - \ref{eq8}\label{halftrans}}
\end{figure}

One should note that equations \ref{eq1} - \ref{eq8} specify a transformation
of five of the six variables amongst themselves. Geometrically we may associate
triangles, for some choice of metric, to each symbol $|abc|$ and can, thus,
show the above equations graphically in figure \ref{halftrans} where the left
hand side is embedded into a space with a Minkowski signature metric and the
edges are regarded as timelike vectors. We shall discuss the geometry  in more
detail in section \ref{geom}.

Eight similar equations may be derived connecting $a,b,d,e,f$ and 
$a^\prime,b^\prime,d^\prime,e^\prime,f^\prime$ to which may be associated a very similar
geometry to figure \ref{halftrans}. Here $f = f^\prime$ is the shared edge.

The left hand side of equations \ref{eq1} - \ref{eq8}, and the analogous
equations connecting  $a,b,d,e,f$ to 
$a^\prime,b^\prime,d^\prime,e^\prime,f^\prime$,
being positive is equivalent to the symbol 
$\left| \begin{array}{ccc}
a & b & c \\ d & e & f \end{array} \right|_{SU(2)}$ being defined, while positivity
of the right hand side is equivalent to the symbol 
$\left| \begin{array}{ccc}
a^\prime & b^\prime & c^\prime \\
d^\prime & e^\prime & f^\prime \end{array} \right|_{SU(1,1)}$ being defined. 
So the map is well defined and by definition the following anti-triangle
inequalities are enforced
\begin{align}
c^\prime \ge & a^\prime+b^\prime+1 \label{tri1} \\
e^\prime \ge & d^\prime+c^\prime+1 \label{tri2} \\
e^\prime \ge & a^\prime+f^\prime+1 \label{tri3} \\
f^\prime \ge & b^\prime+d^\prime+1 \label{tri4} 
\end{align}

We may also define the extension  
$\left\{ \begin{array}{ccc}
a^\prime & b^\prime & c^\prime \\
d^\prime & e^\prime & f^\prime \end{array} \right\}_{ext}$
of the SU(2) 6j symbol to the anti-triangle inequality
domain via the map $S$.

\begin{defi}
\begin{equation}
\left\{ \begin{array}{ccc}
a^\prime & b^\prime & c^\prime \\
d^\prime & e^\prime & f^\prime \end{array} \right\}_{ext}
:= \left\{ \begin{array}{ccc}
a & b & c \\ d & e & f \end{array} \right\}_{SU(2)}
\end{equation}
where 
\begin{multline} \label{6jdef}
\left\{ \begin{array}{ccc}
a & b & c \\ d & e & f \end{array} \right\}_{SU(2)}
= \Delta(abc)\Delta(cde)\Delta(bdf)\Delta(aef)\\ \times
\sum_n \frac{(-1)^n (n+1)!}{(n-a-b-c)!(n-c-d-e)!(n-b-d-f)!(n-a-e-f)!}\\ \times
\frac{1}{(a+b+d+e-n)!(a+c+d+f-n)!(b+c+e+f-n)!}
\end{multline}
where $\Delta(abc) = \sqrt{\frac{(a+b-c)!(a-b+c)!(-a+b+c)!}{(a+b+c+1)!}}$ 

When any of  the factorials are undefined $\left\{ \begin{array}{ccc} a & b & c
\\ d & e & f \end{array} \right\}_{SU(2)}$ is defined to be zero. 
This requirement ensures the sum over $n$ is finite,
restricts the indices to non negative half integers and ensures that $a+b+c$,
etc are always integer. \end{defi}

All symmetries of the `extended' 6j symbol
may be reduced to permutations and sign changes in certain 
variables\cite{Pon74_6}. 
Thus for the 6j symbol $\left\{ \begin{array}{ccc}
a & b & c \\ d & e & f \end{array} \right\}$, we define the variables

\begin{align*}
s_1 & = a+d+1 & s_0 & = d-a \\
s_3 & = b+e+1 & s_2 & = e-b \\
s_5 & = c+f+1 & s_4 & = f-c 
\end{align*}

Then all permutations of the $s_i$, or sign changes of an even number of the
$s_i$, give the total number of extended symmetries of the associated 6j 
symbol. The
Regge symmetries\footnote{By which we mean the 144 symmetries 
that preserve the triangle inequalities}\cite{Reg59} correspond to permutations of $\left(
s_0,s_2,s_4\right)$ or $\left(s_1,s_3,s_5\right)$, and sign changes of any two of
$\left(s_0,s_2,s_4\right)$. 

Let $s^\prime_{\sigma\left(i\right)} = s_i$, then
the symmetry that corresponds to the map $S$ above is simply the following
permutation, $\sigma$,
$$
\sigma =   
\left( \begin{array}{cccccc}
0 & 1 & 2 & 3 & 4 & 5 \\
0 & 2 & 3 & 1 & 4 & 5
\end{array} \right)
$$

and from equations \ref{tri1} - \ref{tri4} it is easy to see the transformation
$S$ takes
us into the region characterised by anti triangle inequalities, conjectured
to be the 6j symbol for the discrete unitary representations of SU(1,1).

For $\left\{ \begin{array}{ccc}
a & b & c \\ d & e & f \end{array} \right\}_{SU(2)}$ we have the well known
orthogonality relation
\begin{equation}
\sum_X\left(2X+1\right)\left\{ \begin{array}{ccc}
a & b & X \\ c & d & p \end{array} \right\}_{SU(2)}\left\{ \begin{array}{ccc}
a & b & X \\ c & d & q \end{array} \right\}_{SU(2)} =\delta_{pq}
\frac{\{apd\}_{SU(2)}\{bcp\}_{SU(2)}}{2p+1}
\end{equation}

Our notation $\{apd\}$ is a `triangular delta function', by which we mean it is
zero when the corresponding symbol $|adp|$ is undefined and one when
the symbol $|apd|$ is defined.

By transforming everything in this equation with $S$, it is easy to see a
similiar relation holds for $\left\{ \begin{array}{ccc}
a & b & c \\ d & e & f \end{array} \right\}_{ext}$. In the latter case,
however, the right hand side will be non zero when anti-triangle inequalities
are satisfied by the relevant three indices. In both cases one has the geometric
interpretation of two tetrahedra, glued together along two common faces, for
the left hand side of the equation.

Thus we may state
\begin{prop}
\begin{equation}
\sum_{X^\prime}\left(2X^\prime+1\right)\left\{ \begin{array}{ccc}
a^\prime & b^\prime & X^\prime \\ c^\prime & d^\prime & p^\prime \end{array} 
\right\}_{ext}\left\{ \begin{array}{ccc}
a^\prime & b^\prime & X^\prime \\ c^\prime & d^\prime & q^\prime \end{array} 
\right\}_{ext} =\delta_{p^\prime q^\prime}
\frac{\{a^\prime p^\prime d^\prime\}_{SU(1,1)}
\{b^\prime c^\prime p^\prime\}_{SU(1,1)}}{2p^\prime+1}
\end{equation}
\end{prop}

The other crucial relation, the Biedenharn-Elliot identity, is far less
straightforward to see and will be proved in section \ref{rac}.

\section{The Racah Coefficient for SU(1,1)} \label{rac}

In order to derive the Biedenharn-Elliot identity for the extension
we shall derive a formula for the 6j symbol of the positive
discrete unitary representation series of SU(1,1) and show explicitly its
relation to the extension of the SU(2) 6j symbol we have defined.

We shall start with two lemmas that will be of use later

\begin{lem} \label{lem1}
$$
\sum_n (-1)^n\frac{(x+n-1)!}{(z-n)!(y+n-1)!n!}=(-1)^z
\frac{(x-1)!(x-y)!}{z!(y+z-1)!(x-y-z)!}
$$
\end{lem}
\begin{lem} \label{lem2}
$$
\sum_n \frac{1}{(x-n)!(y+n-1)!(z-n)!n!} = 
\frac{(x+y+z-1)!}{x!z!(x+y-1)!(y+z-1)!}
$$
\end{lem}

Lemma \ref{lem1} follows from Gauss' formula for summing
the $ _2F_1$ hypergeometric series\cite{HypGeo}
$$
\sum_n \frac{(a+n-1)!(b+n-1)!(c-1)!}{(a-1)!(b-1)!(c+n-1)!n!} =
\frac{(c-a-b-1)!(c-1)!}{(c-a-1)!(c-b-1)!} $$
with $a=x$, $b = -z$, $c= y$.

Lemma \ref{lem2} follows from  
the addition theorem for binomial coefficients by expanding both sides of
$(a+b)^n (a+b)^m = (a+b)^{n+m}$ and equating powers of $a$ and $b$.

The Lie Algebra $\mathfrak{su}(1,1)$ is defined by generators 
$J_z$, $J_+$ and $J_-$ with relations 
\begin{align*}
\left[J_z,J_\pm\right] & = \pm J_\pm\\
\left[J_-,J_+\right] & = 2J_z
\end{align*}

The positive discrete series is characterised by the following action of the
generators on the Hilbert spaces $\mathcal{H}_j$ with basis 
$\{\ket{j,m} |\; j,m\in\frac{1}{2}\N,\;\; m\ge j\}$
\begin{align*}
J_z\ket{j,m} & = m\ket{j,m}\\
J_\pm\ket{j,m} & = \pm\sqrt{(m\pm j)(m\mp j\pm 1)}\ket{j,m\pm 1}
\end{align*}

The Clebsch-Gordon coefficients are defined as follows
\begin{equation} 
\ket{j,m} = \sum_{m_1,m_2} \left[
\begin{array}{ccc} j_1 & j_2 & j\\ m_1 & m_2 & m \end{array} \right]
\ket{j_1,m_1}\otimes\ket{j_2,m_2}
\end{equation}

A specific formula may be derived by adapting the analysis of \cite{LK92} to the
$q=1$ case. It is found to be

\begin{multline} \label{cg}
\left[ \begin{array}{ccc} j_1 & j_2 & j\\ m_1 & m_2 & m \end{array} \right] = 
\delta_{m_1+m_2,m}(-1)^{m_1-j_1}\Delta(j_1 j_2 j)\\ \times
\sqrt{\frac{(2j-1)(m-j)!(m_1-j_1)!(m_2-j_2)!(m_1+j_1-1)!(m_2+j_2-1)!}{(m+j-1)!}}
\\ \times\sum_z
\frac{(-1)^z}{z!(m-j-z)!(m_1-j_1-z)!(m_1+j_1-z-1)!(j-j_2-m_1+z)!
(j+j_2-m_1+z-1)!}
\end{multline}
where $\Delta(j_1 j_2 j) = \sqrt{(j-j_1-j_2)!(j-j_1+j_2-1)!(j+j_1-j_2-1)!
(j+j_1+j_2-2)!}$.

For SU(2) one defines the Racah coefficients via the recoupling identity

\begin{equation} \label{RC}\mbox{\raisebox{-13mm}{ 
\begin{texdraw} \drawdim{mm} \rmove(-1
-4)\htext{$j_1$}\rmove(1 4) \rlvec(8 16) 
\rlvec(8 -16)  
\rmove(-1 -4)\htext{$j_3$}\rmove(1 4) 
\rmove(-8 0) 
\rmove(-1 -4)\htext{$j_2$}\rmove(1 4)
\rlvec(4 8)  
\rmove(0 4)\htext{$j_{23}$}\rmove(0 -4) 
\rmove(-4 8) \rlvec(0 8)
\rmove(0 1)\htext{$j$}\rmove(0 -1) 
\rmove(115 -24) 
\rmove(-1 -4)\htext{$j_1$}\rmove(1 4) 
\rlvec(8 16) \rlvec(8 -16)  
\rmove(-1 -4)\htext{$j_3$}\rmove(1 4) 
\rmove(-8 0) 
\rmove(-1 -4)\htext{$j_2$}\rmove(1 4)
\rlvec(-4 8)  
\rmove(-4 4)\htext{$j_{12}$}\rmove(4 -4) 
\rmove(4 8) \rlvec(0 8)
\rmove(0 1)\htext{$j$}\rmove(0 -1) 
\rmove(-115 -22) 
\htext{\parbox{300pt}{$$\;\;
=\sum_{j_{12}}\;(-1)^{j_1+j_2+j_3+j}\;\sqrt{(2j_{12}+1)(2j_{23}+1)}\;\left\{  
\begin{array}{ccc} j_1 & j_2 & j_{12}\\ 
j_3 & j& j_{23} \end{array}\right\}$$}} 
\end{texdraw} }} 
\end{equation} 

where each
trivalent vertex is a graphical representation of a Clebsch-Gordon coefficient

\begin{equation}
\mbox{\raisebox{-12mm}{\hspace{-5cm}
\begin{texdraw}
\drawdim{mm}
\lvec(5 10) \move(10 0) \lvec(5 10) 
\lvec(5 17)
\move(-1.5 -4) \htext{$j_1$}
\move(8.5 -4) \htext{$j_2$}
\move(5 18) \htext{$j_{12}$}
\move(11 5)
\htext{$\;\equiv\;\left[ \begin{array}{ccc} j_1 & j_2 & j_{12}\\ 
\phi_1 & \phi_2 & \phi_{12} \end{array} \right]$}
\end{texdraw}
}}
\end{equation}

In terms of the Clebsch-Gordon coefficients for SU(2), the recoupling identity
may be written

\begin{multline}
\sum_{j_{12}\; \phi_{12}} (-1)^{j_1+j_2+j_3+j}\sqrt{(2j_{12}+1)(2j_{23}+1)}
\left\{ \begin{array}{ccc} j_1 & j_2 & j_{12}\\ j_3 & j & j_{23} 
\end{array}\right\}_{SU(2)}
\left[ \begin{array}{ccc} j_1 & j_2 & j_{12}\\ 
\phi_1 & \phi_2 & \phi_{12} \end{array} \right]_{SU(2)}\\
\times\left[ \begin{array}{ccc} j_{12} & j_3 & j\\ 
\phi_{12} & \phi_3 & \phi \end{array} \right]_{SU(2)}
= \sum_{\phi_{23}} 
\left[ \begin{array}{ccc} j_2 & j_3 & j_{23}\\
\phi_2 & \phi_3 & \phi_{23} \end{array} \right]_{SU(2)}
\left[ \begin{array}{ccc} j_1 & j_{23} & j\\ 
\phi_1 & \phi_{23} & \phi \end{array} \right]_{SU(2)}
\end{multline}

where $\left[ \begin{array}{ccc} j_1 & j_2 & j_{12}\\  \phi_1 & \phi_2 &
\phi_{12} \end{array} \right]_{SU(2)}$ are the Clebsch-Gordon coefficients for
the coupling of two unitary irreducible representations of SU(2). In analogy
with the SU(2) case, the equivalent relation for the SU(1,1) Clebsh-Gordon
coefficients for the positive discrete series will be taken  as the definition
of the Racah coefficient of this series. However for the SU(1,1) case the factor
$\sqrt{(2j_{12}+1)(2j_{23}+1)}$ will be replaced by a factor 
$\sqrt{(2j_{12}-1)(2j_{23}-1)}$. This to due to the fact that the formula for
the SU(2) Clebsch-Gordon coefficient has a factor $\sqrt{2j+1}$ (see for instance
\cite{AngMom}) whereas that for the SU(1,1) Clebsch-Gordon coefficient has a
factor $\sqrt{2j-1}$ as in equation \ref{cg}.

One should note that the Clebsch-Gordon coefficient in equation \ref{cg} is
normalised in the sense that
\begin{equation}
\sum_{m_1}
\left[ \begin{array}{ccc} j_1 & j_2 & j\\ m_1 & m-m_1 & m \end{array} \right]
\left[ \begin{array}{ccc} j_1 & j_2 & j\\ m_1 & m-m_1 & m \end{array} \right]
 = 1 
\end{equation}

This relation may be used to bring the defining relation for the Racah
coefficient into the following form
\begin{multline} \label{racdef}
\sqrt{(2c-1)(2f-1)}(-1)^{a+b+d+e}
\left\{ \begin{array}{ccc} a & b & c\\ d & e & f \end{array}
\right\}_{SU(1,1)}
\left[ \begin{array}{ccc} a & f & e\\ \alpha & f & \alpha +f \end{array} \right]
\\ = \sum_\beta 
\left[ \begin{array}{ccc} a & b & c\\ \alpha & \beta & \alpha +\beta \end{array} \right]
\left[ \begin{array}{ccc} c & d & e\\ \alpha +\beta & f-\beta & \alpha +f \end{array} \right]
\left[ \begin{array}{ccc} b & d & f\\ \beta & f-\beta & f \end{array} \right]
\end{multline}

Substitution of equation \ref{cg} into \ref{racdef} gives the following

\begin{multline} \label{rac2}
\left\{ \begin{array}{ccc} a & b & c\\ d & e & f \end{array} 
\right\}_{SU(1,1)} = (-1)^{a+b+d+e}
\frac{\Delta(abc)\Delta(bdf)\Delta(cde)}
{\Delta(afe)}\\ \times  (e-a-f)!(e+a-f-1)!(f+\alpha -e)!\mathcal{I}(\alpha)
\end{multline}
where
\begin{multline*}
\mathcal{I}(\alpha) = 
\sum_{\beta\; t\; u}\frac{(-1)^{t+u}(\alpha +\beta -c)!}
{t!(\alpha +\beta -c-t)!(c-b-\alpha +t)!(c+b-\alpha +t-1)!(\alpha -a -t)!
(\alpha +a-t-1)!}\\
\times\frac{1}{u!(\alpha +f-e-u)!(e-d-\alpha -\beta+u)!(e+d-\alpha -\beta+u-1)!}
\\ \times\frac{1}{
(\alpha +\beta -c-u)!(\alpha +\beta +c-u-1)!}
\end{multline*}

$\mathcal{I}(\alpha)$ may be reduced to a single summation as follows. Introduce
two new summation variables, $m$ and $n$, in place of $\beta$ and $u$ such that
\begin{align*}
u & = \alpha +\beta -c -n\\
\beta & = c-\alpha+m+n
\end{align*}
Then
\begin{multline}
\mathcal{I}(\alpha) = 
\sum_{m\; n\; t}\frac{(-1)^{t+m} (m+n)!}{t!(t+m+n)!(c-b-\alpha +t)!(c+b-\alpha
+t-1)!(\alpha -a -t)!(\alpha +a -t-1)!}\\ \times \frac{1}{
m!(f-e+\alpha -m)!(e+d-c-1-n)!n!(2c-1+n)!(e-d-c-n)!}
\end{multline}

The sum over $m$, using lemma \ref{lem1}, is found to be
$$
\sum_m \frac{(-1)^m(m+n)!}{m!(m+n-t)!(f-e+\alpha -m)!}
= \frac{(-1)^{f-e+\alpha} n!t!}{(f-e+\alpha)!(n+f-e+\alpha -t)!(t-f+e-\alpha)!}
$$
and the sum may be written as
\begin{multline} \label{I1}
\mathcal{I}(\alpha) = 
\sum_{n\; t}\frac{(-1)^{f-e+\alpha -t}}
{(c-b-\alpha +t)!(c+b-\alpha +t-1)!(\alpha -a -t)!(\alpha +a-t-1)!(e+d-c-1-n)!}
\\ \times \frac{1}
{(2c-1+n)!(f-e+\alpha)!(n+f-e+\alpha -t)!(t-f+e-\alpha)!(e-d-c-n)!}
\end{multline}
Now, transforming with $n = -c-d+e -s$, we may rewrite equation \ref{I1} as
\begin{multline}
\mathcal{I}(\alpha)=\sum_{t\; s}\frac{(-1)^{f+\alpha-t-e}}
{(c-b-\alpha+t)!(c+b-\alpha +t-1)!(\alpha -a-t)!(\alpha +a-t-1)!(2d-1+s)!}\\
\times\frac{1}{(c-d+e-1-s)!(f-e+\alpha)!(f-c-d-s+\alpha -t)!(t-f+e-\alpha)!s!}
\end{multline}
The sum over $s$, using lemma \ref{lem2}, is found to be
\begin{multline*}
\sum_s\frac{1}{(2d-1+s)!(c-d+e-1-s)!(f-c-d-s+\alpha -t)!s!}\\ = 
\frac{(e+f-2-t+\alpha)!}{(c-d+e-1)!(f-c-d-t+\alpha)!(c+d+e-2)!(f-c+d-t+\alpha
-1)!}
\end{multline*}
and $\mathcal{I}(\alpha)$ is reduced to a single summation
\begin{multline}
\mathcal{I}(\alpha)=\sum_t\frac{(-1)^{\alpha -t+f-e}(e+f-2-t+\alpha)!}
{(c-b-\alpha +t)!(c+b-\alpha +t-1)!(\alpha-a-t)!(\alpha +a-t-1)!(f-e+\alpha)!
}\\ \times
\frac{1}{(t-f+e-\alpha)!(c-d+e-1)!(f-c-d-t+\alpha)!(c+d+e-2)!(f-c+d-t+\alpha -1)!}
\end{multline}
If the summation variable is rewritten as $z=\alpha -a -t$ and substituted into
equation \ref{rac2} we find
\begin{multline} 
\left\{ \begin{array}{ccc} a & b & c\\ d & e & f \end{array} 
\right\}_{SU(1,1)} = 
\frac{(-1)^{f+2a+b+d}\Delta(abc)\Delta(bdf)\Delta(cde)(e-a-f)!(e+a-f-1)!}
{\Delta(afe)(c+d+e-2)!(c-d+e-1)!}\\ \times
\sum_z\frac{(-1)^z (e+f-2+a+z)!}{z!(c-b-a-z)!(c+b-a-1-z)!(e-f-a-z)!}\\ \times
\frac{1}{(2a-1+z)!(f-c-d+a+z)!(f-c+d+a+z-1)!}
\end{multline}
 
The sum may then be brought into the following, more symmetrical, form
\begin{multline} \label{RacahSU11}
\left\{ \begin{array}{ccc} a & b & c\\ d & e & f \end{array} 
\right\}_{SU(1,1)} = 
\frac{(-1)^{a+b+d-e+1}\Delta(abc)\Delta(bdf)\Delta(cde)(e-a-f)!(e+a-f-1)!}
{\Delta(afe)(c+d+e-2)!(c-d+e-1)!}\\ \times
\sum_r\frac{(-1)^r (r+1)!}{(c-b+e+f-3-r)!(c+b+e+f-4-r)!(2e-3-r)!}\\ \times
\frac{1}{(r+a-e-f+2)!(r-a-e-f+3)!(r+d-c-e+3)!(r-a-e-f+3)!}
\end{multline}

It is of interest to establish the relationship between the Racah coefficients
$\left\{ \begin{array}{ccc} a & b & c\\ d & e & f \end{array} 
\right\}_{SU(1,1)}$ and the SU(1,1) region of the extended Racah
coefficients  $\left\{ \begin{array}{ccc} a & b & c\\ d & e & f \end{array} 
\right\}_{ext}$

\begin{thm} \label{rrelation}
\begin{equation} 
\left\{ \begin{array}{ccc} a+1 & b+1 & c+1\\ d+1 & e+1 & f+1 \end{array} 
\right\}_{SU(1,1)} = (-1)^{a+b+d-e+1}\left\{ \begin{array}{ccc} a & b & c\\
 d & e & f \end{array} 
\right\}_{ext}
\end{equation}
\end{thm}

The proof is simply to transform Racah's form for the SU(2) 6j symbol (see,
for instance, equation \ref{6jdef}) via the transformation $S^{-1}$ given by equations
\ref{a}-\ref{f} and compare that to equation \ref{RacahSU11}. This settles the 
claim of D'Adda, D'Auria and Ponzano, in \cite{Pon74_6}, that the extension of
the SU(2) Racah coefficient was related to the SU(1,1) Racah coefficient and
demonstrates the exact relationship.

Since $\left\{ \begin{array}{ccc} a & b & c\\ d & e & f \end{array} 
\right\}_{SU(1,1)}$ is the associator for the monoidal category of unitary
positive discrete representations, it automatically satisfies the
Biedenharn-Elliot identity in view of the Pentagon relation for associators of
monoidal categories. Theorem \ref{rrelation} implies the SU(1,1) region of the
extended 6j symbol also satisfies a  Biedenharn-Elliot type relation. 

\begin{figure}
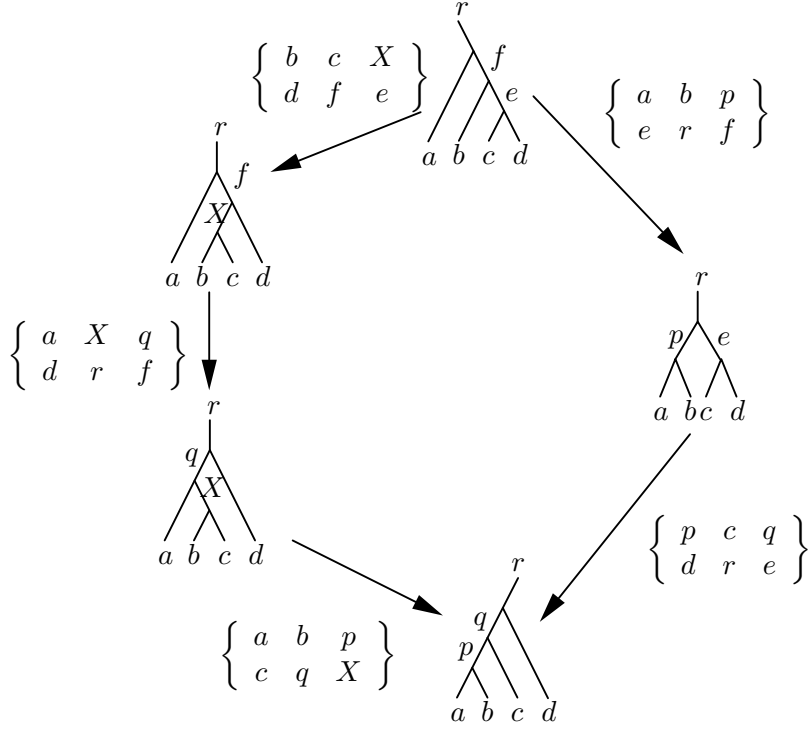

\hspace{2cm}
\begin{texdraw}

\drawdim{mm}

\rmove(-1 -3)\htext{$d$}\rmove(1 3)
\rlvec(-2 4) \rlvec(-2 -4) 
\rmove(-1 -3)\htext{$c$}\rmove(1 3)
\rmove(2 4) \rlvec(-2 4) 
\rmove(2 -2.5)\htext{$e$}\rmove(-2 2.5)
\rlvec(-4 -8) 
\rmove(-1 -3)\htext{$b$}\rmove(1 3)
\rmove(4 8)
\rmove(0 1.5)\htext{$f$}\rmove(0 -1.5)
\rlvec(-2 4) \rlvec(-6 -12) 
\rmove(-1 -3)\htext{$a$}\rmove(1 3)
\rmove(6 12)
\rlvec(-2 4)
\rmove(-0.5 1)\htext{$r$}\rmove(0.5 1) 

\rmove(10 -12)
\arrowheadtype t:F 
\ravec(20 -21) \rmove(-20 21) 
\rmove(-10 12)

\rmove(19 -19) 
\htext{$\left\{
\begin{array}{ccc}
a & b & p \\
e & r & f
\end{array}\right\}$}
\rmove(-19 19)

\rmove(37 -52)
\rmove(-1 -3)\htext{$d$}\rmove(1 3)
\rlvec(-2 5) \rlvec(-2 -5) 
\rmove(-1 -3)\htext{$c$}\rmove(1 3)
\rmove(-2 0)
\rmove(-1 -3)\htext{$b$}\rmove(1 3)
\rlvec(-2 5) \rlvec(-2 -5) 
\rmove(-1 -3)\htext{$a$}\rmove(1 3)
\rmove(2 5)
\rmove(-1 1.5)\htext{$p$}\rmove(1 -1.5)
\rlvec(3 5) \rlvec(3 -5) 
\rmove(-0.5 2)\htext{$e$}\rmove(0.5 -2)
\rmove(-3 5)
\rlvec(0 4) 
\rmove(-0.5 1)\htext{$r$}\rmove(0.5 -1)
\rmove(-2 6)

\rmove(1 -25)
\arrowheadtype t:F 
\ravec(-20 -25) \rmove(20 25) 
\rmove(-1 15)

\rmove(-5 -35) 
\htext{$\left\{
\begin{array}{ccc}
p & c & q \\
d & r & e
\end{array}\right\}$}
\rmove(5 35)

\rmove(-30 -50)
\rmove(-1 -3)\htext{$a$}\rmove(1 3)
\rlvec(2 4) \rlvec(2 -4) 
\rmove(-1 -3)\htext{$b$}\rmove(1 3)
\rmove(-2 4)
\rlvec(2 4) 
\rmove(-4 -3)\htext{$p$}\rmove(4 3)
\rlvec(4 -8) 
\rmove(-1 -3)\htext{$c$}\rmove(1 3)
\rmove(-4 8)
\rmove(-2 1)\htext{$q$}\rmove(2 -1)
\rlvec(2 4) \rlvec(6 -12) 
\rmove(-1 -3)\htext{$d$}\rmove(1 3)
\rmove(-6 12)
\rlvec(2 4) 
\rmove(-1 1)\htext{$r$}\rmove(1 -1)

\rmove(-30 5)
\arrowheadtype t:F 
\ravec(20 -10) \rmove(-20 10) 
\rmove(30 -5)

\rmove(-40 -15) 
\htext{$\left\{
\begin{array}{ccc}
a & b & p \\
c & q & X
\end{array}\right\}$}
\rmove(40 15)

\rmove(-35 5)
\rmove(-1 -3)\htext{$d$}\rmove(1 3)
\rlvec(-6 12) \rlvec(0 4) 
\rmove(-0.5 1)\htext{$r$}\rmove(0.5 -1)
\rmove(0 -4)
\rmove(-3.5 -2)\htext{$q$}\rmove(3.5 2)
\rlvec(-6 -12) 
\rmove(-1 -3)\htext{$a$}\rmove(1 3)
\rmove(4 8)
\rmove(0.5 -2)\htext{$X$}\rmove(-0.5 2)
\rlvec(4 -8) 
\rmove(-1 -3)\htext{$c$}\rmove(1 3)
\rmove(-2 4)
\rlvec(-2 -4)
\rmove(-1 -3)\htext{$b$}\rmove(1 3)

\rmove(2 33)
\arrowheadtype t:F 
\ravec(0 -13) \rmove(0 13) 
\rmove(-2 -33)

\rmove(-25 20) 
\htext{$\left\{
\begin{array}{ccc}
a & X & q \\
d & r & f
\end{array}\right\}$}
\rmove(25 -20)

\rmove(-3 37)
\rmove(-1 -3)\htext{$a$}\rmove(1 3)
\rlvec(6 12) \rlvec(0 4) 
\rmove(-0.5 1)\htext{$r$}\rmove(0.5 -1)
\rmove(0 -4)
\rmove(2 -2)\htext{$f$}\rmove(-2 2)
\rlvec(6 -12) 
\rmove(-1 -3)\htext{$d$}\rmove(1 3)
\rmove(-4 8)
\rmove(-4 -2.5)\htext{$X$}\rmove(4 2.5)
\rlvec(-4 -8) 
\rmove(-1 -3)\htext{$b$}\rmove(1 3)
\rmove(2 4)
\rlvec(2 -4)
\rmove(-1 -3)\htext{$c$}\rmove(1 3)

\rmove(25 20)
\arrowheadtype t:F 
\ravec(-20 -8) \rmove(20 8) 
\rmove(-25 -20)

\rmove(2 20) 
\htext{$\left\{
\begin{array}{ccc}
b & c & X \\
d & f & e
\end{array}\right\}$}
\rmove(-2 -20)
\end{texdraw}

\caption{The Pentagon relation \label{pent}}
\end{figure}

The Pentagon relation for the associator of a monoidal category is shown in
figure \ref{pent}. It asserts the equivalence of the two ways of moving from
$(V_a\otimes (V_b \otimes (V_c \otimes V_d)))$ to 
$(((V_a\otimes V_b) \otimes V_c) \otimes V_d)$ so that the diagram is
commutative.

One may read off the Biedenharn-Elliot relation  for the SU(1,1) Racah
coefficients from figure \ref{pent}. Once the appropriate normalisation and
phase, from the SU(1,1) version of the graphical recoupling relation in
equation \ref{RC}, is inserted for each 6j symbol the following may be derived

\begin{prop}[Biedenharn-Elliot relation for SU(1,1)]
\begin{multline}\sum_X(-1)^R\left(2X-1\right)
\left\{\begin{array}{ccc}
b & c & X \\d & f & e
\end{array}\right\}_{SU(1,1)}
\left\{\begin{array}{ccc}
a & X & q \\d & r & f
\end{array}\right\}_{SU(1,1)}
\left\{\begin{array}{ccc}
a & b & p \\c & q & X
\end{array}\right\}_{SU(1,1)} = \\
\left\{\begin{array}{ccc}
a & b & p \\e & r & f
\end{array}\right\}_{SU(1,1)}
\left\{\begin{array}{ccc}
p & c & q \\d & r & e
\end{array}\right\}_{SU(1,1)}
\end{multline}
where $R = a+b+c+d+e+f+p+q+r+X$ 
\end{prop}

If one adopts the same geometric interpretation of 6j symbols being tetrahedra,
as in the SU(2) case, then this equation has the geometric interpretation of
three tetrahedra glued along a common edge (labelled by $X$) being transformed
into two tetrahedra glued along a common face (labelled by $e$, $r$ and $p$).
The exact geometric interpretation of each SU(1,1) 6j symbol is discussed in
section \ref{geom}.

\section{Geometry} \label{geom}

We wish to explore the geometry of the extended 6j symbols for the SU(1,1)
region. It is known (see
\cite{PR},\cite{BaFo93}) that the symbol
$\left| \begin{array}{ccc}
a & b & c \\ d & e & f \end{array} \right|_{SU(2)}$
may be  identified with a Euclidean, or spacelike
Lorentzian, tetrahedron with edge
lengths equal to $j_{12} = a+\frac{1}{2}$, etc. Here a spacelike Lorentzian
tetrahedra is one for which all faces and all edges are spacelike.
We shall denote such a tetrahedron by
$T\left(j_{12},j_{13},j_{14},j_{34},j_{24},j_{23}\right)$, and omit the edge
lengths when these are not relevant. We shall also use subscripts, SU(2) and
SU(1,1), to indicate the region the tetrahedron is associated to when confusion
can arise. Note that we shall impose the requirement that the edge lengths 
in the symbol $T$ be positive for the SU(1,1) case\footnote{
While $j_{12},j_{13}$, etc. are always positive for $T_{SU(2)}$ 
the same cannot be said for 
$T_{SU(1,1)}$.
An easy counter example is gained by mapping a regular tetrahedron to the
SU(1,1) domain with equations \ref{a} - \ref{f}. So this assumption is
necessary.}.

\begin{figure}[htb!]
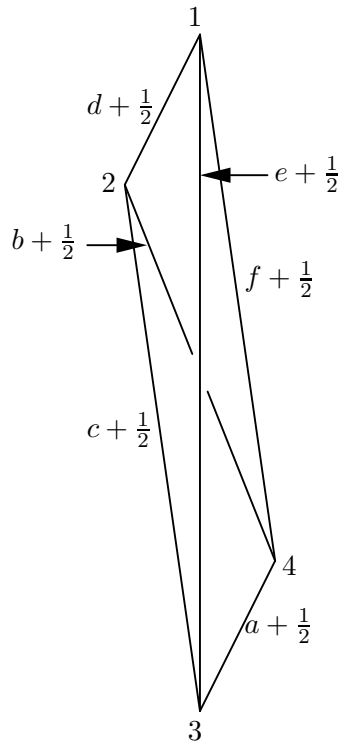

\begin{center}
\begin{texdraw}
\drawdim{mm}
\lvec(10 20) 
\move(6 9)\htext{$a+\frac{1}{2}$} 
\move(0 0) \lvec(0 90)
\move(10 69) \htext{$e+\frac{1}{2}$}
\move(9 71.3)
\arrowheadtype t:F \avec(0 71.3)
\move(0 0) \lvec(-10 70)
\move(-15 35) \htext{$c+\frac{1}{2}$}
\move(10 20) \lvec(0 90)
\move(6 55)  \htext{$f+\frac{1}{2}$}
\move(10 20) \lvec(1 42.5)
\move(-1 47.5) \lvec(-10 70)
\move(-25 60) \htext{$b+\frac{1}{2}$}
\move(-15 62)
\arrowheadtype t:F \avec(-7 62)
\move(-10 70) \lvec(0 90)
\move(-15 78)  \htext{$d+\frac{1}{2}$}
\move(-1.5 -4)\htext{3}
\move(-1.5 91)\htext{1}
\move(-13 69)\htext{2}
\move(11 18)\htext{4}
\end{texdraw}
\end{center}
\caption{A Lorentzian tetrahedron with all edges and all faces timelike.
Time increases vertically up the page.\label{tltri}}
\end{figure}

To fix notation we shall denote the length of the edge $(h,k)$, 
formed by deleting the $h$-$th$ and $k$-$th$ vertex
(see figure \ref{tltri}), as $j_{hk}$.
The area $A_h$ denotes the area of the
face, $\mathcal{T}_{h}$,
obtained by deleting the $h$-th vertex from the tetrahedron. It is clear we may
associate a geometric triangle, $\mathcal{T}$, to each symbol $|abc|$.

We shall denote by $\theta_{hk}$
the (exterior) dihedral angle on the edge $(h,k)$
between the two outward normals of the faces 
$\mathcal{T}_h$ and $\mathcal{T}_k$. In Euclidean space these are always 
bone fide real
angles; for Lorentzian space the situation is more subtle since the `angles' can
turn out to be complex. This situation has been analysed in some detail in
\cite{BaFo93} and we shall say more about this in section \ref{Cayneg}.

Associated to each $T$ is a number, $V^2$, given by the
Cayley determinant which defines the volume squared of the tetrahedron.
\begin{equation}
V^2 = \frac{1}{2^3\left(3!\right)^2}
\left|\begin{array}{ccccc}
0 & j_{34}^2 & j_{24}^2 & j_{23}^2 & 1\\
j_{34}^2 & 0 & j_{14}^2 & j_{13}^2 & 1\\
j_{24}^2 & j_{14}^2 & 0 & j_{12}^2 & 1\\
j_{23}^2 & j_{13}^2 & j_{12}^2 & 0 & 1\\
1 & 1 & 1 & 1 & 0\\
\end{array}\right|
\end{equation}

$T_{SU(2)}$ is
Euclidean if, and only if, the Cayley
determinant is positive and Minkowskian when
it is negative. For edge lengths that are positive half
integers the Cayley determinant cannot vanish.

For $T_{SU(1,1)}$ we claim it may be identified with a  tetrahedron whose faces
are timelike and edges are either all spacelike or all timelike. These timelike
triangles have one `long' side and two `short' sides. As such they obey
anti-triangle inequalities along the lines of $$ c \ge a + b $$ where c is the
`long' side. The normals to such triangles are spacelike, and  the triangles
possess two interior `angles', which are complex and may thus be identified
with Lorentzian boosts as in \cite{BaFo93} (opposite the `a' and `b' sides),
with the third interior angle being undefined\footnote{If the edges are
timelike this interior angle would involve boosting from the future light cone
to the past light cone, which can't be done. If the edges are spacelike it
involves boosting through either the past, or future, light cone.}, and one
exterior `angle' (for the vertex opposite the `c' side) which may, again, be
identified with a  Lorentzian boost.  The area squared defined by $A^2 = 
\frac{1}{16}\left(a+b+c\right)\left(a+b-c\right)\left(a-b+c\right)\left(-a+b+c\right)$
is negative. The area, as in the triangle inequality case, may be defined by
taking the square root of the area squared, so that
$A=i\sqrt{\left|A^2\right|}$.

Equations \ref{tri1} - \ref{tri4} specify how to fit four such
timelike triangles together. 
The resulting object has one
`super long' edge ($j_{24}$), two `long' edges ($j_{14}$ and $j_{23}$) and
the remaining three are `short' edges. An embedding of such an object into
Minkowski space is shown in figure \ref{tltri}. 

Figure \ref{tltri} is the general form for such a tetrahedron. If the edges are
timelike  there must be a strict time ordering (up to time reversal) of the
vertices.  Once we have choosen such an ordering  (say 1,2,4,3\footnote{Our
choice of numbering comes from attempting to preserve conventions with
\cite{PR}}  from future to past) the `super long' edge connects vertex 1 to
vertex 3, the two long edges connect vertex 1 to vertex 4 and vertex 2 to
vertex 3, and the remaining vertices are connected by short edges.

One should note that if the symbol $\left| \begin{array}{ccc} a & b & c \\ d &
e & f \end{array} \right|_{SU(2)}$  has a `degenerate' triangle (ie $a+b=c$ for
some triangle $|abc|$) then the corresponding tetrahedron, $T_{SU(2)}$ has an
`almost degenerate' triangle, that is  $j_{12} + j_{13}= j_{14}+\frac{1}{2}$.
The +1's in equations \ref{tri1} - \ref{tri4} ensure the same is true for the
SU(1,1) case.

We now state a proposition relating 
$T_{SU(1,1)}$ and
$T_{SU(2)}$.

\begin{prop} \label{p1}
Let $T\left(j_{12},j_{13},j_{14},j_{34},j_{24},j_{23}\right)_{SU(2)}$ and 
$T\left(j_{12}^\prime,j_{13}^\prime,j_{14}^\prime,
j_{34}^\prime,j_{24}^\prime,j_{23}^\prime\right)_{SU(1,1)}$
be related by equations \ref{a} - \ref{f}. 
 
Then the transformation preserves the Cayley determinant
and the product of the associated face areas.
\end{prop}
\textbf{Proof}
\textit{
Straightforward, if laborious, algebra. 
}

There are two geometric cases to consider depending on whether 
the Cayley determinant is positive or negative.

\subsection{The case where $V^2 > 0$} \label{Caypos}

If the Cayley determinant
is positive we choose an embedding of $T_{SU(1,1)}$
in Lorentzian space with metric signature
$(+,-,-)$ so that the timelike edges have a positive length squared. 
Moreover, since the
normals to the faces span a spacelike plane, all the dihedral angles are 
defined, in contrast to the spacelike case discussed in \cite{BaFo93}.

We now wish to consider how the dihedral angles of the tetrahedra transform under
equations \ref{b1} - \ref{b6} in this case. 
In contrast to the Regge symmetries the sum of dihedral
angles times edge lengths does not remain constant.

\begin{thm} \label{dihedral_pos}
Under equations \ref{b1} - \ref{b6} the dihedral angles transform as:
\begin{align}
\theta_{12} & = \pi-\frac{1}{2}\left(\theta_{12}^\prime+\theta_{13}^\prime-
\theta_{34}^\prime+\theta_{24}^\prime\right) \label{a1}\\
\theta_{13} & = -\frac{1}{2}\left(-\theta_{12}^\prime-\theta_{13}^\prime-
\theta_{34}^\prime+\theta_{24}^\prime\right)\\
\theta_{14} & = \pi - \theta_{14}^\prime \\
\theta_{34} & = \pi - \frac{1}{2}\left(-\theta_{12}^\prime+\theta_{13}^\prime+
\theta_{34}^\prime+\theta_{24}^\prime\right) \\
\theta_{24} & = 2\pi - \frac{1}{2}\left(\theta_{12}^\prime-\theta_{13}^\prime+
\theta_{34}^\prime+\theta_{24}^\prime\right)\\
\theta_{23} & = \pi -\theta_{23}^\prime \label{a2}
\end{align}
for $V^2>0$.
\end{thm}

The proof involves the following Euclidean 
trigonometric relations between dihedral angles 
and edge lengths:

\begin{align} 
-C_{rs} = 16 A_r A_s\cos\theta_{rs} & \;\;\;  & r\not= s \label{cos}\\
\frac{3}{2}Vj_{rs} = A_r A_s\sin\theta_{rs} & & r\not= s \label{sin}
\end{align}

where $j_{rs}$ is the shared side for the triangles whose areas are given by 
$A_r$ and $A_s$, $\theta_{rs}$ is the (exterior) dihedral angle
between the outward normals to the faces $\mathcal{T}_r$ and $\mathcal{T}_s$,
and $C_{rs}$ is the $(r,s)$
algebraic minor of the Cayley determinant formed by deleting the row and
the column common to the $(r,s)$ matrix entry. Note that equation \ref{sin} does not 
distinguish exterior and interior dihedral angles, whereas equation \ref{cos} does.

To derive equation \ref{sin} for the Lorentzian case one must choose a 
square root of the identity
\begin{equation}
V^2 = \frac{4 A_h^2 A_k^2}{9j_{hk}^2}\sin^2\theta_{hk} 
\;\;\;\;\;\;\;\;\;\;\;\;\;\;\;\;\;\;  h\not= k
\end{equation}
so that the dihedral angle has the correct range, that is 
$0\le\theta_{hk}\le\pi$.
Thus, since $\left(A_h^\prime\right)^2 < 0$, we must choose
\begin{equation}
V = \frac{2 |A^\prime_h| |A^\prime_k|}{3j^\prime_{hk}}\sin\theta^\prime_{hk}
\;\;\;\;\;\;\;\;\;\;\;\;\;\;\;\;\;\;  h\not= k\label{sin_lor}
\end{equation}

Now, since we want to use the fact that, from 
proposition \ref{p1}, 
\begin{equation}
 A_1 A_2 A_3 A_4 = A^\prime_1 A^\prime_2 A^\prime_3 A^\prime_4 = 
|A^\prime_1| |A^\prime_2| |A^\prime_3| |A^\prime_4|
\end{equation}
in the following proof,
we must rewrite equation \ref{cos} in a 
similar way. Thus, for $T_{SU(1,1)}$
\begin{align}
-C_{rs}^\prime = & 16 A^\prime_r A^\prime_s\cos\theta^\prime_{rs}\nonumber\\
= &  -16 |A^\prime_r| |A^\prime_s|\cos\theta^\prime_{rs}\nonumber\\
= & 16 |A^\prime_r| |A^\prime_s|\cos\left(\pi - \theta^\prime_{rs}\right)
   \label{cos_lor}
\end{align}

Note that equation \ref{cos_lor} now gives \emph{interior} dihedral angles.
In the following we shall use the Euclidean formulae, equations \ref{cos} and \ref{sin},
for $T_{SU(2)}$ on the left hand side of the following equations 
and the Lorentzian formulae,
equations \ref{sin_lor} and \ref{cos_lor}, for 
$T_{SU(1,1)}$ on the right side of the following equations,
thus we get interior rather than exterior angles for the SU(1,1)
case. To prevent confusion we shall denote an interior dihedral angle as
$\bar{\theta}_{hk}$ and so we have $\pi -\bar{\theta}_{hk} = \theta_{hk}$

We may show
\begin{align}
\sin\left(\theta_{12}+\theta_{34}\right) = &
\sin\left(\bar{\theta}_{13}^\prime+\bar{\theta}_{24}^\prime\right)\label{ex}\\
\sin\left(\theta_{12}-\theta_{34}\right) = &
\sin\left(\bar{\theta}_{12}^\prime-\bar{\theta}_{34}^\prime\right)\\
\sin\left(\theta_{13}+\theta_{24}\right) = &
\sin\left(\bar{\theta}_{24}^\prime-\bar{\theta}_{13}^\prime\right)\\
\sin\left(\theta_{13}-\theta_{24}\right) = &
\sin\left(-\bar{\theta}_{12}^\prime-\bar{\theta}_{34}^\prime\right)\\
\sin\theta_{14} = & \sin\bar{\theta}_{14}^\prime\\
\sin\theta_{23} = & \sin\bar{\theta}_{23}^\prime\label{ex1}
\end{align}

The proof is simple, if laborious, algebra; for instance, by using equations 
\ref{cos}, \ref{sin}, \ref{sin_lor}, \ref{cos_lor}
and proposition \ref{p1}, equation \ref{ex} may be reduced to
showing
\begin{equation}
j_{12}C_{34}+j_{34}C_{12} = j_{13}^\prime C_{24}^\prime + j_{24}^\prime C_{13}^\prime
\end{equation}
which follows directly from algebra.

The same equations, with sines replaced by cosines, may be drived 
in a similar way; so we conclude, since all the 
$\theta_{ij}, \theta_{ij}^\prime\in\left[0,\pi\right]$,

\begin{align}
\theta_{12}+\theta_{34} & = \bar{\theta}_{13}^\prime+\bar{\theta}_{24}^\prime \label{di1}\\
\theta_{12}-\theta_{34} & = \bar{\theta}_{12}^\prime-\bar{\theta}_{34}^\prime \label{di2}\\
\theta_{13}+\theta_{24} & = \bar{\theta}_{24}^\prime-\bar{\theta}_{13}^\prime + 2n_1\pi\label{di3}\\
\theta_{13}-\theta_{24} & = -\bar{\theta}_{12}^\prime-\bar{\theta}_{34}^\prime + 2n_2\pi\label{di4}\\
\theta_{14} & = \bar{\theta}_{14}^\prime\\
\theta_{23} & = \bar{\theta}_{23}^\prime
\end{align}

where the $n_i = 1$ or $0$.

And hence that
\begin{align}
\theta_{12} & = \pi - \frac{1}{2}\left(\theta_{12}^\prime+\theta_{13}^\prime-
\theta_{34}^\prime+\theta_{24}^\prime\right)\label{ang1}\\
\theta_{13} & = -\pi-\frac{1}{2}\left(-\theta_{12}^\prime-\theta_{13}^\prime-
\theta_{34}^\prime+\theta_{24}^\prime\right)+\left(n_1+n_2\right)\pi \\
\theta_{14} & = \pi-\theta_{14}^\prime\label{ang2}\\
\theta_{34} & = \pi-\frac{1}{2}\left(-\theta_{12}^\prime+\theta_{13}^\prime+
\theta_{34}^\prime+\theta_{24}^\prime\right)\\
\theta_{24} & = \pi-\frac{1}{2}\left(\theta_{12}^\prime-\theta_{13}^\prime+
\theta_{34}^\prime+\theta_{24}^\prime\right)+\left(n_1-n_2\right)\pi\label{ang3}\\
\theta_{23} & = \pi-\theta_{23}^\prime\label{ang6}
\end{align}
where we are now relating the \emph{exterior} dihedral angles.

Now, the sum of the interior dihedral angles around any 
vertex for a Euclidean
tetrahedron are greater than $\pi$, while those for the top and bottom vertices of the 
SU(1,1) tetrahedron are less
than $\pi$. Indeed for every vertex of a Euclidean tetrahedron
one may associate a spherical
triangle whose interior angles correspond to the tetrahedron's interior
dihedral angles; each
of the three triangles meeting at a given vertex defines a plane and the intersection
of these planes with a sphere defines the triangle. For a $T_{SU(1,1)}$ the top and
bottom vertices define hyperbolic triangles via an intersection with hyperbolic space
in much the same way.

Thus, from equations \ref{ang1}, \ref{ang2} and \ref{ang3},
\begin{equation} \label{te1}
2\pi > \left(\theta_{12}+\theta_{24}+\theta_{14}\right)
 = 3\pi - \left(\theta_{12}^\prime+\theta_{24}^\prime+\theta_{14}^\prime\right)
+\left(n_1-n_2\right)\pi
\end{equation}
where $$\theta_{12}^\prime+\theta_{24}^\prime+\theta_{14}^\prime > 2\pi$$

Now consider a long thin $T_{SU(1,1)}$ that is on the verge of degenerating into
a line. We have
$j_{14}^\prime+j_{34}^\prime\approx j_{24}^\prime
\approx j_{12}^\prime+j_{23}^\prime$
with
$\theta_{12}^\prime+\theta_{24}^\prime+\theta_{14}^\prime\approx 2\pi$
which implies for $T_{SU(2)}$ 
$j_{12}+j_{13}\approx j_{14}$ and $j_{13}+j_{34}\approx j_{23}$ so that
$\theta_{12}+\theta_{24}+\theta_{14}\approx 2\pi$

Thus, in this case, we have $n_1 = 1$ and $n_2 = 0$. Now vary the edge lengths
$j_{hk}^\prime$ continuously. Since the dihedral angles depend continuously 
on the edge lengths,
the angles will vary continuously between 0 and $\pi$. 
Thus, by continuity, the
result holds generally; which concludes the proof of 
theorem \ref{dihedral_pos}.

\subsection{The case where $V^2 < 0$} \label{Cayneg}

If the Cayley determinant is negative then we do not have the above embedding
into Minkowski space. It is clear the metric has signature $(+,+,-)$ or
$(-,-,-)$, but the latter, being equivalent to an embedding into Euclidean
space, cannot happen. Thus geometrically we embed in a spacetime with metric
$(+,+,-)$ and regard the edges of the tetrahedron as spacelike, while the faces
must still be timelike since they satisfy anti-triangle inequalities.

If we define the dihedral angles in the same way to the previous discussion
then, in both cases, they are complex. These complex angles will be called
exterior or interior depending on whether the defining equation gave exterior
or interior dihedral angles in section \ref{Caypos}.

The possible Lorentzian boosts that take the place of the dihedral angles in
this case come in two flavours, either an  interior boost is defined with no
possible exterior boost, or vice versa.  Since the normals to the faces and the
edges are spacelike, the normals span a plane in  Minkowski space and there
will be no exterior boost defined  when two normals are separated by the
lightcone. A similar criterion determines the existence of interior boosts.

There are only two patterns that may occur. Either one has three interior
boosts, around one face, with the remainder exterior. Here opposite edges have
different flavours of boost. Or, one has two exterior boosts and four interior
boosts, with opposite edges having the same flavour. This should be compared to
the spacelike Lorentzian case for $T_{SU(2)}$\cite{BaFo93} where an identical
situation arises for analogous reasons. In the following the first case will be
referred to as a \emph{type 1} tetrahedron and the second as a \emph{type 2}
tetrahedron for both the $T_{SU(2)}$ and $T_{SU(1,1)}$ cases.

We use the following conventions in making sense of these complex dihedral
angles\cite{BaFo93} that arise when one tries to use the Euclidean formula to
define the dihedral angles. 

For
$T_{SU(2)}$ we choose an embedding into Lorentzian spacetime with
metric $(-,+,+)$ (so that the sign of the Cayley determinant is preserved by the
transformation). Thus an interior dihedral boost is given by
$$ \Theta_{hk} = \cosh^{-1}\left(n_h\cdot n_k\right) $$
while an exterior dihedral boost is given by 
$$ \Theta_{hk} = -\cosh^{-1}\left(-n_h\cdot n_k\right) $$

where $n_i$ is the outward normal to the $i$-th triangle.  In the first case
the complex angle $\theta$, given by the usual Euclidean formula, has the form
$\theta_{hk} = \pi +  i\Im\theta_{hk}$, while for the second it is pure
imaginary.

For $T_{SU(1,1)}$
we embed into a spacetime as above. Here an exterior dihedral boost
is given by
$$ \Theta_{hk}^\prime = -\cosh^{-1}\left(n_h^\prime\cdot n_k^\prime\right) $$
while the interior dihedral boost is given by
$$ \Theta_{hk}^\prime = \cosh^{-1}\left(-n_h^\prime\cdot n_k^\prime\right) $$
since the normals are spacelike and $n^2 = 1$ for a spacelike unit vector $n$.
Similarly we have $\theta_{hk}^\prime$ as pure imaginary for exterior angles,
while $\theta_{hk}^\prime = \pi + i\Im\theta_{hk}^\prime$ for interior angles.

In view of this we make the obvious identification $\Theta_{hk} =
\Im\theta_{hk}$, where $\Theta_{hk}$ is a Lorentzian boost. Such a boost is an
interior dihedral boost when it arises as the imaginary part of a complex angle
given by the usual Euclidean formula for interior angles. Otherwise it will be
called an exterior dihedral boost.

We now state and prove a theorem about the transformation of these Lorentzian
boosts.
\begin{thm} \label{dihedral_neg}
Under equations \ref{b1} - \ref{b6} the boosts transform as:
\begin{align}
\Theta_{12} & = -\frac{1}{2}\left(\Theta_{12}^\prime+\Theta_{13}^\prime-
\Theta_{34}^\prime+\Theta_{24}^\prime\right)\\
\Theta_{13} & = -\frac{1}{2}\left(-\Theta_{12}^\prime-\Theta_{13}^\prime-
\Theta_{34}^\prime+\Theta_{24}^\prime\right)\\
\Theta_{14} & = -\Theta_{14}^\prime \\
\Theta_{34} & = -\frac{1}{2}\left(-\Theta_{12}^\prime+\Theta_{13}^\prime+
\Theta_{34}^\prime+\Theta_{24}^\prime\right) \\
\Theta_{24} & = -\frac{1}{2}\left(\Theta_{12}^\prime-\Theta_{13}^\prime+
\Theta_{34}^\prime+\Theta_{24}^\prime\right)\\
\Theta_{23} & = -\Theta_{23}^\prime
\end{align}
for $V^2<0$.
\end{thm}

Our starting point will be the following equations relating complex exterior
angles on the left to complex interior angles on the right, as in the previous
case with the complex angles still given by the normal Euclidean formula

\begin{align}
\sin\left(\theta_{12}+\theta_{34}\right) = &
\sin\left(\bar{\theta}_{13}^\prime+\bar{\theta}_{24}^\prime\right)\label{z1}\\
\sin\left(\theta_{12}-\theta_{34}\right) = &
\sin\left(\bar{\theta}_{12}^\prime-\bar{\theta}_{34}^\prime\right)\\
\sin\left(\theta_{13}+\theta_{24}\right) = &
\sin\left(\bar{\theta}_{24}^\prime-\bar{\theta}_{13}^\prime\right)\\
\sin\left(\theta_{13}-\theta_{24}\right) = &
\sin\left(-\bar{\theta}_{12}^\prime-\bar{\theta}_{34}^\prime\right)\\
\sin\left(\theta_{14}\right) = &
\sin\left(\bar{\theta}_{14}^\prime\right)\\
\sin\left(\theta_{23}\right) = &
\sin\left(\bar{\theta}_{23}^\prime\right)\label{z2}
\end{align}

As before, the same equations with sine replaced by cosine are also valid.
These follow from algebra using the expressions for the sine and cosine of
dihedral angles as in section \ref{Caypos}. We may then expand these using the
standard trignometric formula for angle sums and discard the real part of
equations \ref{z1} - \ref{z2}  (which is clearly identically zero for both
sides). 

Hence we are left with the following:
\begin{align}
\cos\left(\Re\theta_{12}+\Re\theta_{34}\right)
\sinh\left(\Im\theta_{12}+\Im\theta_{34}\right) = &
\cos\left(\Re\bar{\theta}_{13}^\prime+\Re\bar{\theta}_{24}^\prime\right)
\sinh\left(\Im\bar{\theta}_{13}^\prime+\Im\bar{\theta}_{24}^\prime\right)\label{dih1}\\
\cos\left(\Re\theta_{12}-\Re\theta_{34}\right)
\sinh\left(\Im\theta_{12}-\Im\theta_{34}\right) = &
\cos\left(\Re\bar{\theta}_{12}^\prime-\Re\bar{\theta}_{34}^\prime\right)
\sinh\left(\Im\bar{\theta}_{12}^\prime-\Im\bar{\theta}_{34}^\prime\right)\label{dih2}\\
\cos\left(\Re\theta_{13}+\Re\theta_{24}\right)
\sinh\left(\Im\theta_{13}+\Im\theta_{24}\right) = &
\cos\left(\Re\bar{\theta}_{24}^\prime-\Re\bar{\theta}_{13}^\prime\right)
\sinh\left(\Im\bar{\theta}_{24}^\prime-\Im\bar{\theta}_{13}^\prime\right)\\
\cos\left(\Re\theta_{13}-\Re\theta_{24}\right)
\sinh\left(\Im\theta_{13}-\Im\theta_{24}\right) = &
\cos\left(-\Re\bar{\theta}_{12}^\prime-\Re\bar{\theta}_{34}^\prime\right)
\sinh\left(-\Im\bar{\theta}_{12}^\prime-\Im\bar{\theta}_{34}^\prime\right)\\
\cos\left(\Re\theta_{14}\right)
\sinh\left(\Im\theta_{14}\right) = &
\cos\left(\Re\bar{\theta}_{14}^\prime\right)
\sinh\left(\Im\bar{\theta}_{14}^\prime\right)\\
\cos\left(\Re\theta_{23}\right)
\sinh\left(\Im\theta_{23}\right) = &
\cos\left(\Re\bar{\theta}_{23}^\prime\right)
\sinh\left(\Im\bar{\theta}_{23}^\prime\right)
\end{align}

We also gain the same equations with sinh replaced by cosh by taking the real
part of the cosine versions of equations \ref{z1} - \ref{z2}. It is clear, in
the second case, that the result of the cosine must have the same sign for each
side of the equations. From which we can deduce that the tetrahedron type is
preserved by the  transformation and derive (once we have replaced the interior
complex angles on the right handside by exterior complex angles)

\begin{align}
\Im\theta_{12} & = -\frac{1}{2}\left(\Im\theta_{12}^\prime+\Im\theta_{13}^\prime-
\Im\theta_{34}^\prime+\Im\theta_{24}^\prime\right)\\
\Im\theta_{13} & = -\frac{1}{2}\left(-\Im\theta_{12}^\prime-\Im\theta_{13}^\prime-
\Im\theta_{34}^\prime+\Im\theta_{24}^\prime\right)\\
\Im\theta_{14} & = -\Im\theta_{14}^\prime\\
\Im\theta_{34} & = -\frac{1}{2}\left(-\Im\theta_{12}^\prime+\Im\theta_{13}^\prime+
\Im\theta_{34}^\prime+\Im\theta_{24}^\prime\right)\\
\Im\theta_{24} & = -\frac{1}{2}\left(\Im\theta_{12}^\prime-\Im\theta_{13}^\prime+
\Im\theta_{34}^\prime+\Im\theta_{24}^\prime\right)\\
\Im\theta_{23} & = -\Im\theta_{23}^\prime 
\end{align}
which concludes the proof of theorem \ref{dihedral_neg}.

For the transformation of the real part of the complex dihedral angle (as
defined by the Euclidean formula) we have the following result

\begin{thm} \label{real}
Under equations \ref{b1} - \ref{b6} the real parts of the dihedral `angles' 
transform as:
\begin{align}
\Re\theta_{12} & = \pi - \frac{1}{2}\left(\Re\theta_{12}^\prime+\Re\theta_{13}^\prime-
\Re\theta_{34}^\prime+\Re\theta_{24}^\prime\right)\\
\Re\theta_{13} & = -\frac{1}{2}\left(-\Re\theta_{12}^\prime-\Re\theta_{13}^\prime-
\Re\theta_{34}^\prime+\Re\theta_{24}^\prime\right)\\
\Re\theta_{14} & = \pi - \Re\theta_{14}^\prime \\
\Re\theta_{34} & = \pi - \frac{1}{2}\left(-\Re\theta_{12}^\prime+\Re\theta_{13}^\prime+
\Re\theta_{34}^\prime+\Re\theta_{24}^\prime\right) \\
\Re\theta_{24} & = 2\pi-\frac{1}{2}\left(\Re\theta_{12}^\prime-\Re\theta_{13}^\prime+
\Re\theta_{34}^\prime+\Re\theta_{24}^\prime\right)\\
\Re\theta_{23} & = \pi - \Re\theta_{23}^\prime
\end{align}
for $V^2<0$.
\end{thm}

Indeed it is almost obvious that the real parts must transform in the same way
as the dihedral angles for the tetrahedera with positive Cayley determinant. We
argue as follows, the real parts of the angles correspond to a least degenerate
geometric configuration of the edges for an embedding into the space in which
we may legitimately embed the associated positive Cayley determinant tetrahedra.

Thus for $T_{SU(2)}$, type 1 tetrahedra are characterised in Euclidean space by
three of the faces lying flat on one face and failing to meet at a vertex. It is
clear that rotating the faces upwards in Euclidean space simply makes the
configuration more degenerate. Thus the Euclidean `dihedral angles' \emph{are}
given by the real part. The type 2 tetrahedra in this case consist of a pair of
triangles lying flat on another pair of triangles in a least degenerate
configuration as well. Again we find the Euclidean `dihedral angles' given by
the real part.

For $T_{SU(1,1)}$ we have an analogous situation. For instance a type 1
tetrahedron embedded into $(+,-,-)$ Lorentzian space consists of three
overlapping faces lying flat on one face. It is clear that boosting the faces
outwards makes them more degenerate since they overlap more.
Thus we may apply theorem \ref{dihedral_pos} to the real parts of the dihedral
angles by regarding it as simply a transformation of two degenerate positive
Cayley determinant tetrahedra to gain theorem \ref{real} as a corollary.

\section{Asymptotics} \label{asymp}

It is of interest to see if one can find a similar asymptotic formula to the
Ponzano-Regge formula for the SU(2) 6j symbol. Their formula for $V^2 >0$,
from \cite{PR}, is
\begin{equation} 
\left\{ \begin{array}{ccc}
a & b & c \\ d & e & f \end{array} \right\}\sim 
\frac{1}{\sqrt{12\pi V}}\cos\left(\sum^4_{h,k=4}
j_{hk}\theta_{hk}+\frac{\pi}{4}\right) \label{PR_pos1}
\end{equation}

\begin{figure}[htb]
\epsfbox{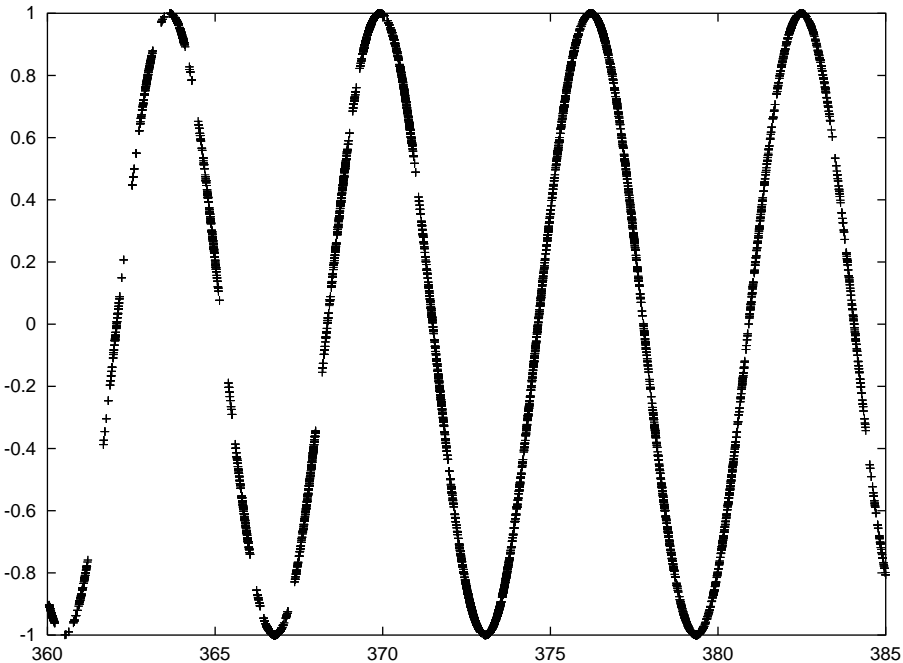}
\caption{
A plot of $\sum^4_{h,k=0}
j_{hk}\theta_{hk}$ (x-axis) versus $\sqrt{12\pi V} \{6j\}$ (y-axis) 
\label{su2}
}
\end{figure}

where $\theta_{hk}$ is defined as previously and $V$ is the volume. There has
never been a direct proof of the validity of this formula but a formula
asymptotic to equation \ref{PR_pos1} has been proven in \cite{Flude,Rob98}  and
numerical results give a good indication of its validity. Indeed we have
plotted some values in figure \ref{su2}, which gives a clear cosine shape.

For the SU(1,1) extension we have been considering, one may, subject to the
validity of equation \ref{PR_pos1}, derive the following
\begin{prop}
\begin{equation}
\left\{ \begin{array}{ccc}
a & b & c \\ d & e & f \end{array} \right\}  \sim\frac{1}{\sqrt{12\pi V}}
\left(-1\right)^{j^\prime_{12} +j_{14}^\prime +j_{34}^\prime+ 2j^\prime_{24}
+j_{23}^\prime}\cos\left(\sum^4_{h,k=4}
j_{hk}^\prime\theta_{hk}^\prime
-\frac{\pi}{4}\right) 
\end{equation}
for $V^2>0$
\end{prop}

In view of theorem \ref{dihedral_pos},
 one should consider how the quantity $\sum^4_{h,k=0}
j_{hk}\theta_{hk}$ transforms under equations \ref{a} - \ref{f}. Using 
equations \ref{a1} - \ref{a2} and the orthogonality of the transformation from 
$T_{SU(2)}$
to $T_{SU(1,1)}$ given by equations \ref{a} - \ref{f}, it is easy to
show that, for $V^2>0$,

\begin{equation}
\sum^4_{h,k=0}j_{hk}\theta_{hk} = -\sum^4_{h,k=0}j_{hk}^\prime \theta_{hk}^\prime
+\left(j^\prime_{12} +j_{14}^\prime +j_{34}^\prime+ 2j^\prime_{24}
+j_{23}^\prime\right)\pi
\end{equation}

\begin{figure}[htb]
\epsfbox{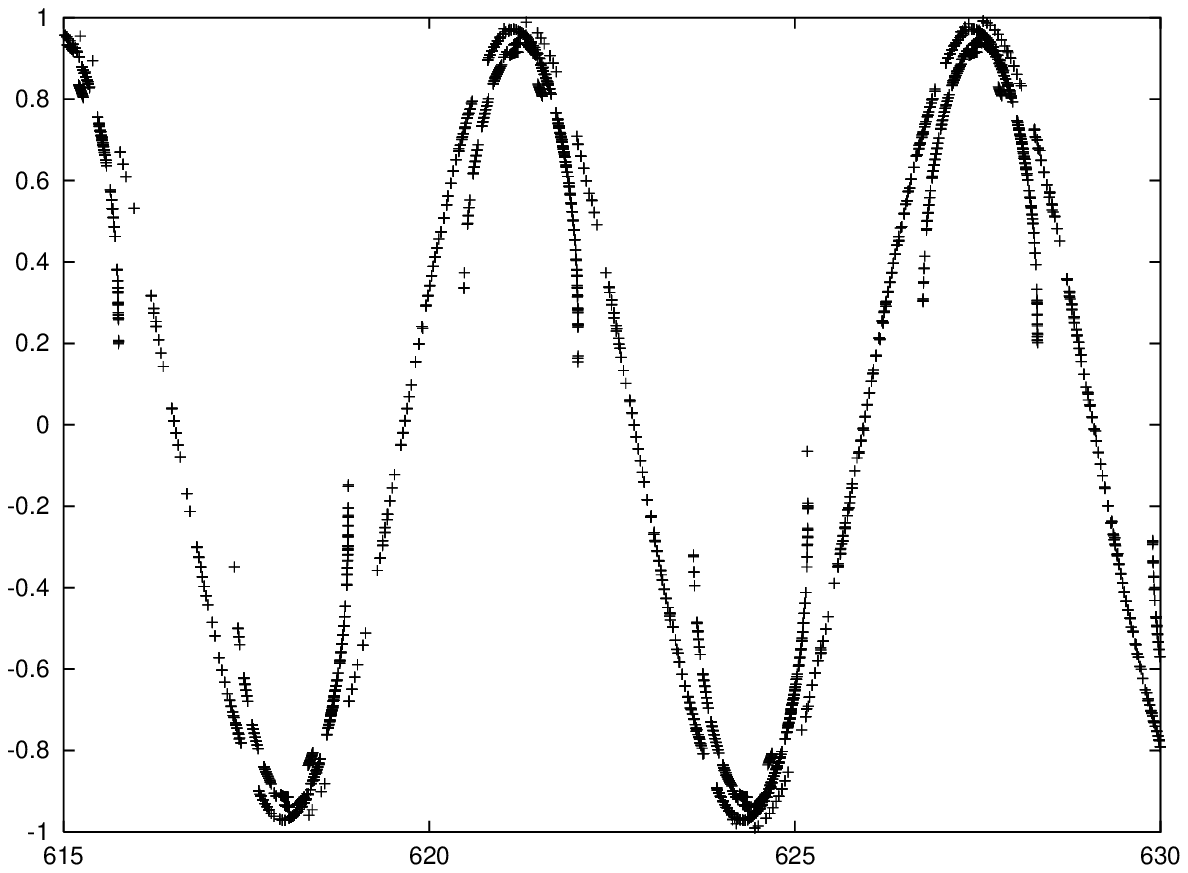}
\caption{
A plot of $-\sum^4_{h,k=0}j_{hk}^\prime \theta_{hk}^\prime
+\left(j^\prime_{12} +j_{14}^\prime +j_{34}^\prime+ 2j^\prime_{24}
+j_{23}^\prime\right)\pi$
 (x-axis) versus $\sqrt{12\pi V}\{6j\}$ (y-axis)
\label{sl2}}
\end{figure}

This, since opposite edge lengths in the tetrahedron always sum to integers,
completes the proof.

We show the validity of this result in figure \ref{sl2}. One might be concerned
by the regions that fall off more steeply than a cosine in the figure,
however numerical results indicate that the tetrahedra in these regions have at
least one face that is
reasonably close to being degenerate, and as such we might expect the above
asymptotic formula to be a worse approximation here. 

One should note the phase factor in front of the cosine.
As mentioned previously, the analgous transformation for the SU(2) 3j
symbol only gives the 3j symbol for SU(1,1) up to a phase factor. Thus we would
expect something similar to happen for the transformation of the 6j symbols.

For the case $V^2 <0$ Ponzano and Regge's exponentially
decaying asymptotic formula for the SU(2) 6j symbol is:

\begin{equation} \label{PR_neg}
\left\{ \begin{array}{ccc}
a & b & c \\ d & e & f \end{array} \right\}\sim 
\frac{1}{2\sqrt{12\pi\left| V\right|}}\cos\phi\exp\left(-\left|\sum^4_{h,k=0}
j_{hk}\mathrm{Im}\theta_{hk}\right|\right)
\end{equation}

where 
\begin{equation}
\cos\phi = \left(-1\right)^{\sum\left(j_{hk}-\frac{1}{2}\right)m_{hk}}
\end{equation}
and $m_{hk}$ is 1 if $\theta_{hk}$ is an interior angle, and 0 otherwise.

There has been no proof of the validity of this formula, although
numerical results provide substantial agreement. Assuming its validity we may
derive the following for the SU(1,1) extension

\begin{prop}
\begin{equation}
\left\{ \begin{array}{ccc}
a & b & c \\ d & e & f \end{array} \right\}  
\sim\frac{1}{2\sqrt{12\pi\left| V\right|}}
\cos\phi^\prime\exp\left(-\left|\sum^4_{h,k=0}
j_{hk}^\prime\Theta_{hk}^\prime\right|\right)
\end{equation}
for $V^2<0$ and $\phi^\prime$ as in equation \ref{phiprime}. 
\end{prop}

Applying theorem \ref{dihedral_neg} and using the orthogonality up to sign
of the transformation as before we see

\begin{equation}
\left|\sum^4_{h,k=0}j_{hk}\Theta_{hk}\right|
 = \left|\sum^4_{h,k=0}j_{hk}^\prime \Theta_{hk}^\prime\right|
\end{equation}

\begin{figure}[htb]
\epsfbox{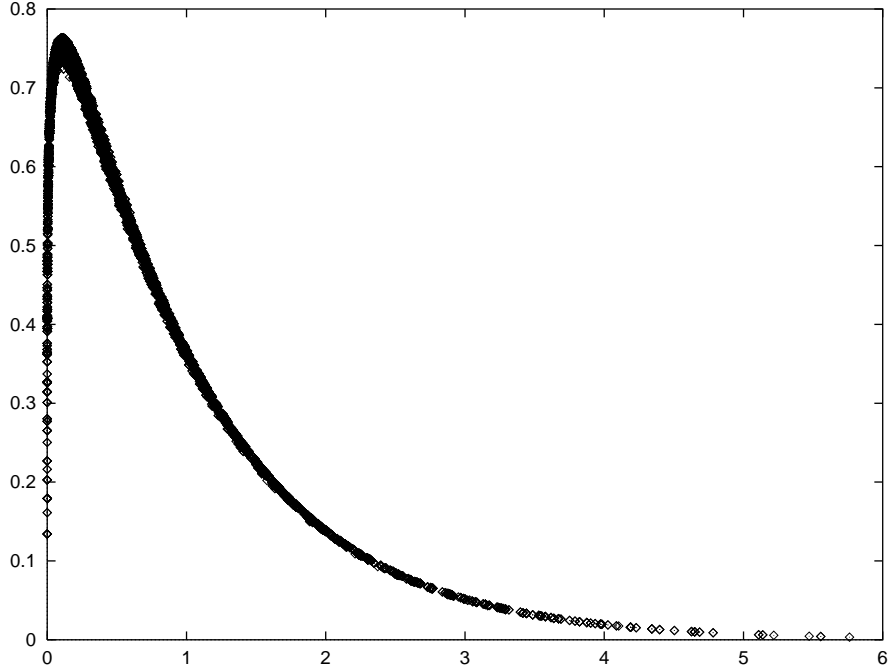}
\caption{
A plot of $\left|\sum^4_{h,k=0}
j_{hk}^\prime\theta_{hk}^\prime\right|$ 
(x-axis) versus $\frac{2\sqrt{12\pi\left|V\right|}}{\cos\phi^\prime}\{6j\}$ (y-axis)
\label{sl2.exp}}
\end{figure}

For the quantity
\begin{equation}
\phi = \sum\left(j_{hk}-\frac{1}{2}\right)\Re\theta_{hk}
\end{equation}
we may use theorem \ref{real} for the transformation
of the real part and apply equations \ref{a} -
\ref{f} to the edge lengths. The resulting transformations are orthogonal
up to a shift depending on the edge lengths and we may derive the following
\begin{equation} \label{phiprime}
\phi^\prime = -\sum_{hk}\left(j_{hk}^\prime-\sigma_{hk}\right)\Re\theta_{hk}^\prime
+(j^\prime_{12} +j_{14}^\prime +j_{34}^\prime+ 2j^\prime_{24}
+j_{23}^\prime - 3)\pi
\end{equation}
where 
\begin{equation} 
\sigma_{hk} = \left\{ \begin{array}
{r@{\quad\mathrm{for}\quad}l} 
0 & (h,k) = (1,2),(1,3),(3,4)\\
-\frac{1}{2} & (h,k) = (1,4),(2,3)\\
-1 & (h,k) = (2,4)
\end{array}\right.
\end{equation}

We have plotted some values for this in figure \ref{sl2.exp} to show the
validity of this result.

\section*{Acknowledgments}

The author wishes to thank John W Barrett for many enlightening
discussions and meticulous reading of draft versions. The author was supported by
an EPSRC research studentship.

\end{document}